\documentclass[fleqn,twoside]{article}
\usepackage{espcrc2}
\usepackage{graphicx}
\usepackage[figuresright]{rotating}

\newcommand{\nubar}{\bar{\nu}}
\newcommand{\numu}{\nu_{\mu}}
\newcommand{\numubar}{\bar{\nu}_{\mu}}
\newcommand{\nue}{\nu_e}
\newcommand{\nuebar}{\bar{\nu}_e}
\newcommand{\mutoe}{\numu\rightarrow\nue}
\newcommand{\mubartoebar}{\numubar\rightarrow\nuebar}
\def\np#1#2#3   {{ Nucl. Phys.} {\bf#1}, #2 (#3)}
\def\pl#1#2#3   {{ Phys. Lett.} {\bf#1}, #2 (#3)}
\def\prev#1#2#3 {{ Phys. Rev.} {\bf#1}, #2 (#3)}

\title{Prospects for Antineutrino Running at MiniBooNE}

\author{M.O. Wascko\address{Department of Physics and Astronomy,\\
                            Louisiana State University, Baton Rouge, LA 70803}}
       
\begin{document}

\begin{abstract}
  We outline a program of antineutrino cross-section measurements
  necessary for the next generation of neutrino oscillation
  experiments, that can be performed with one year of data at
  MiniBooNE.  We describe three independent methods of constraining
  wrong-sign (neutrino) backgrounds in an antineutrino beam, and
  their application to the MiniBooNE antineutrino cross section
  measurements.
\vspace{1pc}
\end{abstract}

\maketitle

\section{Introduction}

The search for CP violation in the neutrino sector requires both
$\mutoe$ and $\mubartoebar$ oscillations measuremetns by future
off-axis experiments.  The signature for CP violation is an asymmetry
in these oscillation probabilities, but this can only be confirmed if the
precision of the $\nu$ and $\nubar$ cross sections are smaller than
the observed asymmetry.  There are few $\nu$ cross section data
published~\cite{xsec-nu} to date, but even fewer measurements of low
energy $\nubar$ cross sections.  We will need more and better data if
we hope to find CP violation.

Table~\ref{table:nubar-event-stats} lists the expected antineutrino
event statistics for one year of $\nubar$ running ($2\times10^{20}$
POT) with MiniBooNE~\cite{ion,loi,nuance}.  Rates are listed for both
right-sign (antineutrino,RS) and wrong-sign (neutrino,WS)
interactions. Note that wrong-sign comprise $30\%$ of the total
events.  To constrain the wrong-sign backgrounds, MiniBooNE has
developed new analysis techniques. We describe three methods below,
and also describe their application to $\nubar$ cross section
measurements at MiniBooNE.

\begin{table}[h]
   \caption{\em Event rates expected in MiniBooNE $\nubar$ running with 
            $2\times10^{20}$ POT assuming a 550 cm fiducial volume,  
            before cuts. Listed are the expected right-sign (RS)
            and wrong-sign (WS) events for each reaction channel.
            These event estimates do not include the 
            effects of final state interactions in carbon 
            nor the effects from event reconstruction.}
   \label{table:nubar-event-stats}
\centering
   \begin{tabular}{ c c c }
\\Reaction & $\numubar$ (RS)  & $\numu$ (WS)  \\ \hline
  CC QE                    &  32,476 &  11,234  \\ 
  NC elastic               &  13,329 &   4,653  \\ 
  CC resonant $1\pi^-$     &   7,413 &      0   \\ 
  CC resonant $1\pi^+$     &       0 &   6,998  \\ 
  CC resonant $1\pi^0$     &   2,329 &   1,380  \\ 
  NC resonant $1\pi^0 $    &   3,781 &   1,758 \\ 
  NC resonant $1\pi^+$     &   1,414 &     654  \\ 
  NC resonant $1\pi^-$     &   1,012 &     520  \\ 
  NC coherent $1\pi^0$     &   2,718 &     438  \\ 
  CC coherent $1\pi^-$     &   4,487 &       0    \\ 
  CC coherent $1\pi^+$     &       0 &     748  \\ 
  other (multi-$\pi$, DIS) &   2,589 &   2,156  \\ \hline
  total                    &  71,547 &  30,539 \\ \hline   
   \end{tabular}
\end{table}

\section{Constraining Wrong Sign Events}
\label{sec:ws-constraints}

For charged current (CC) interactions, neutrino events are typically
distinguished from antineutrino events by identifying the charge of
the outgoing muon.  MiniBooNE, which has no magnetic field, has
developed several novel techniques for measuring wrong-sign
backgrounds in antineutrino mode data, allowing more precise
antineutrino cross section measurements. The wrong-sign content is
constrained by three measurements: muon angular distributions in
quasi-elastic (CC QE) events, muon lifetimes, and the measured rate of
CC single pion (CC1$\pi^+$) events~\cite{loi}.

\subsection{Muon Angular Distributions}

The most powerful wrong-sign constraint comes from the observed
direction of outgoing muons in CC QE interactions. Neutrino and
antineutrino events exhibit distinct muon angular distributions.  Due
to the antineutrino helicity, the final state muons in $\numubar$ QE
events are more forward peaked than muons from $\numu$ interactions.

MiniBooNE's angular resolution allows exploitation of this difference
by fitting the angular distributions to extract the wrong-sign
contribution.  Analysis of Monte Carlo data sets determined the
accuracy with which the wrong-sign content can be measured using this
technique to be $5\%$ of itself~\cite{hiro-memo}.  Including
systematic uncertainties and (non-QE) backgrounds increases the
uncertainty only to $7\%$.

\subsection{Muon Lifetimes}

A second constraint results from measuring the rate at which muons
decay in the MiniBooNE detector. Due to an $8\%$ $\mu^-$ capture
probability in mineral oil, positively and negatively charged muons
exhibit different effective lifetimes ($\tau = 2.026 \, \mu s$ for
$\mu^-$~\cite{mu-lifetime-cap} and $\tau = 2.197 \, \mu s$ for
$\mu^+$~\cite{mu-lifetime-std}).  For CCQE events, we find
that the wrong-sign contribution can be extracted with a $30\%$
statistical uncertainty based solely on this lifetime difference and
negligible systematic uncertainties. While not as precise as fits to
the muon angular distributions, this particular constraint is unique,
as it is independent of kinematics.

\begin{table}
   \caption{\em Wrong-sign extraction uncertainties as obtained from 
            various independent sources in the $\nubar$ data. The
            resultant systematic uncertainty on $\nubar$ cross section
            measurements is obtained by assuming that wrong-signs
            comprise $30\%$ of the total events.}
   \label{table:ws-uncertainties}
\centering
   \begin{tabular}{ c c c }
\\Measurement  &   \hspace{0.2cm}WS \hspace{0.2cm} &  \hspace{0.2cm}resultant \\
                           &   \hspace{0.2cm}uncertainty\hspace{0.2cm} & \hspace{0.2cm} error on $\sigma_{\nubar}$ 
\\ \hline
  CC QE $\cos\theta_\mu$   &  $7\%$   &  $2\%$   \\
  CC $1\pi^+$ cuts           &  $15\%$  &  $5\%$   \\
  muon lifetimes                     &  $30\%$  &  $9\%$   \\ \hline
   \end{tabular}
\end{table}

\subsection{CC Single Pion Event Sample}

Our third wrong-sign constraint employs the the fact that
antineutrinos do not create any CC$1\pi^+$ events in the
detector---these all stem from neutrinos
(Table~\ref{table:nubar-event-stats}). MiniBooNE identifies CC$1\pi^+$
events by tagging the two decay electrons that follow the primary
neutrino interaction, one from the $\mu^-$ and one from the $\pi^+$
decay~\cite{morgan-ccpip}.  However, CC$1\pi^-$ events do not pass
this requirement because most of the emitted $\pi^-$'s are absorbed in
carbon, leaving no decay electrons. Applying these cuts to the full
sample, which is $70\%$ antineutrino (RS) interactions, yields an
$85\%$ pure sample of WS neutrino events.

Assuming conservative uncertainties for the antineutrino background
events and the CC$1\pi^+$ cross section, which is currently being
measured by MiniBooNE, we expect a $15\%$ uncertainty on the
wrong-sign content in the beam given $2\times10^{20}$ POT.  This
constraint is complementary to the muon angular distributions because
CC$1\pi^+$ events stem mainly from resonance decays, thus constraining
the wrong-sign content at larger neutrino energies.

\subsection{Summary of Wrong Sign Constraints}
\label{subsec:ws-summary}

0 The three separate techniques to measure the wrong-sign
content in the antineutrino data will lend confidence to
the antineutrino cross section measurements and greatly reduce their
associated systematics. Combined, these three independent measurements
(each of which have different systematics) offer a very powerful
constraint on the neutrino backgrounds in antineutrino mode
(Table~\ref{table:ws-uncertainties}).

\section{CC Quasi-Elastic Scattering}

MiniBooNE expects more than 40,000 QE interactions in antineutrino
mode with $2\times10^{20}$ POT before cuts. Using the same QE event
selection criteria as the previously reported MiniBooNE neutrino
analysis~\cite{jocelyn-ccqe} yields a sample of $\sim19,000$ events,
with $75\%$ ($\numu + \numubar$) QE purity.

Assuming the wrong-sign constraint from
Section~\ref{sec:ws-constraints} and conservative errors on the $\nu$
flux, the backgrounds, and event detection, we expect a MiniBooNE
measurement of the antineutrino QE cross section to better than $20\%$
with $2\times10^{20}$ POT.

\section{NC Single Pion Production}

There has been only one published measurement of the absolute rate of
$\numubar$ NC $\pi^0$ production, with $25\%$ uncertainty at 2
GeV~\cite{ncpi0-faissner}. This channel is the largest background to
future $\numubar\!\rightarrow\!\nuebar$ oscillation searches.

Applying MiniBooNE's $\numu$ NC $\pi^0$ cuts~\cite{jen-pi0}, with no
modifications, leaves a sample of antineutrino NC $\pi^0$ events with
a similar event purity and efficiency. After this selection, we expect
1,650 $\numubar$ resonant NC $\pi^0$ events and 1,640 $\numubar$
coherent NC $\pi^0$ events assuming $2\times10^{20}$
POT~\cite{nuance,rein-sehgal-coh}. The WS background of $\sim$~1000 WS
events will be known from the constraints on the wrong-sign content in
the beam as described in Section~\ref{sec:ws-constraints} and the
measurement of the $\numu$ NC $\pi^0$ cross section from MiniBooNE
neutrino data.

\section{CC Single Pion(CC1$\pi^-$) Production}

MiniBooNE expects roughly 7,000 resonant CC $1\pi^-$ with
$2\times10^{20}$ POT before cuts. As discussed above, most of the
emitted $\pi^-$'s will be absorbed by carbon nuclei, and will
therefore not be selected by the CC1$\pi^+$ cuts. Nevertheless, these
events still have a signature: two Cherenkov rings (one each from the
$\mu^+$ and $\pi^-$) and one Michel electron in the vicinity of the
$\mu^-$. The selection efficiency and purity of such events is unknown
at this time.  Further investigation is currently underway.

\section{Conclusions}

We have developed three techniques for determining the wrong-sign
background in antineutrino mode.  The resulting systematic error on
any $\nubar$ cross section measurement due to the wrong sign
contamination is around $2\%$, with a total uncertainty around 20\%,
which is remarkable for a detector which does not possess
event-by-event sign selection. Given this redundant approach, the
wrong-sign contamination should not be considered prohibitive to
producing meaningful antineutrino cross section~\cite{loi} and
oscillation measurements~\cite{loi,jocelyn-memo,alexis-memo} at
MiniBooNE. These techniques may also be useful for other experiments
without magnetized detectors which have plans to study antineutrino
interactions ({\em e.g.} T2K, NO$\nu$A, Super-K).

\section{Acknowledgments}

The author is pleased to acknowledge the collaborative efforts of
J.M. Link, H.A. Tanka, and G.P. Zeller in developing the ideas in this
work.

The MiniBooNE collaboration gratefully acknowledges support from
various grants and contracts from the Department of Energy and the
National Science Foundation. The author was supported by grant number
DE-FG02-91ER0617 from the Department of Energy.


\begin{thebibliography}{9}

\bibitem{xsec-nu} G.~P.~Zeller, NuInt02,  {\em hep-ex/0312061}.

\bibitem{ion}I.~Stancu, these proceedings.

\bibitem{loi} MiniBooNE Collaboration, {\em ``Addendum to the MiniBooNE Run Plan: MiniBooNE Physics in 2006,''}
available from {\texttt http://www-boone.fnal.gov/publicpages/loi.ps.gz}.

\bibitem{nuance} D.~Casper, Nucl. Phys. Proc. Suppl. {\bf 112}, 161
                 (2002), {\em hep-ph/0208030}.

\bibitem{hiro-memo} H.A.~Tanaka, ``Estimating Wrong Sign Contamination in Negative Polarity Horn Data'', BooNE Memo (2004)

\bibitem{mu-lifetime-cap} T.Suzuki {\em et al.}, \prev{C35}{2122}{1987} .

\bibitem{mu-lifetime-std} S.Eidelman {\em et al.}, \pl{B592}{33}{2004} .

\bibitem{morgan-ccpip} M.~O.~Wascko, DPF04,  {\em hep-ex/0412008}.

\bibitem{jocelyn-ccqe} J.~Monroe,  Moriond 2004, {\em hep-ex/0406048}.

\bibitem{ncpi0-faissner} H.Faissner,{\em et al.}, \pl{126B}{230}{1983} .

\bibitem{jen-pi0} J.~L.~Raaf, NuInt04, {\em hep-ex/0408015}.

\bibitem{rein-sehgal-coh} D.~Rein and L.~M.~Sehgal, \np{B223}{29}{1983} .

\bibitem{jocelyn-memo} J.~Monroe,  "$\nu_{\mu}$ Disappearance Studies for the Fall 2004 Letter of Intent", BooNE Memo (2004)

\bibitem{alexis-memo} A.A.~Aguilar~Arevalo, "MiniBooNE Oscillation Sensitivity in Antineutrino Running Mode", BooNE Memo (2004)

\end{thebibliography}
\end{document}